\documentclass[letter]{aa} 
\pdfoutput=1 
\usepackage{amsmath,amstext}
\usepackage[T1]{fontenc}
\usepackage{txfonts}
\usepackage{graphicx,subcaption}
\usepackage{color}
\usepackage{multirow}
\usepackage{rotating}
\usepackage{pdflscape}
\usepackage{float}
\usepackage{appendix}

\usepackage{natbib,twoopt}
\usepackage[breaklinks=true]{hyperref} 
\bibpunct{(}{)}{;}{a}{}{,}             
\makeatletter
  \newcommandtwoopt{\citeads}[3][][]{\href{http://adsabs.harvard.edu/abs/#3}%
    {\def\hyper@linkstart##1##2{}%
     \let\hyper@linkend\@empty\citealp[#1][#2]{#3}}}
  \newcommandtwoopt{\citepads}[3][][]{\href{http://adsabs.harvard.edu/abs/#3}%
    {\def\hyper@linkstart##1##2{}%
     \let\hyper@linkend\@empty\citep[#1][#2]{#3}}}
  \newcommandtwoopt{\citetads}[3][][]{\href{http://adsabs.harvard.edu/abs/#3}%
    {\def\hyper@linkstart##1##2{}%
     \let\hyper@linkend\@empty\citet[#1][#2]{#3}}}
  \newcommandtwoopt{\citeyearads}[3][][]%
    {\href{http://adsabs.harvard.edu/abs/#3}
    {\def\hyper@linkstart##1##2{}%
     \let\hyper@linkend\@empty\citeyear[#1][#2]{#3}}}
\makeatother

\begin{document}

\title{X-ray quasi-periodic eruptions from the galactic nucleus of\\ RX J1301.9+2747}

   \author{Margherita Giustini\inst{1}
          \and
          Giovanni Miniutti\inst{1}
          \and
          Richard D. Saxton\inst{2}}

   \institute{Centro de Astrobiolog\'ia (CSIC-INTA), Camino Bajo del Castillo s/n, Villanueva de la Ca\~nada, E-28692 Madrid, Spain\\
   \and
   Telespazio-Vega UK for ESA, Operations Department;
European Space Astronomy Centre (ESAC), Camino Bajo del Castillo s/n,
Villanueva de la Ca\~nada, E-28692 Madrid, Spain        }

   \date{Received  / Accepted  }

\abstract
   {
Following the recent discovery of X-ray quasi-periodic eruptions (QPEs) coming from the nucleus of the galaxy GSN 069, here we report on the detection of QPEs in the active galaxy named RX J1301.9+2747.
QPEs are rapid and recurrent increases of the X-ray count-rate by more than one order of magnitude with respect to a stable quiescent level.
During a XMM-Newton observation lasting 48 ks   that was performed on 30 and 31 May 2019, three strong QPEs lasting about half an hour each were detected in the light curves of RX J1301.9+2747.
The first two QPEs are separated by a longer recurrence time (about 20 ks) compared to the second and third (about 13 ks). This pattern is consistent with the alternating long-short recurrence times of the GSN 069 QPEs, although the difference between the consecutive recurrence times is significantly smaller in GSN 069.
Longer X-ray observations will better clarify the temporal pattern of the QPEs in RX J1301.9+2747 and will allow a detailed comparison with GSN 069 to be performed.
The X-ray spectral properties of QPEs in the two sources are remarkably similar, with QPEs representing fast transitions from a relatively cold and likely disk-dominated state to a state that is characterized by a warmer emission similar to the so-called soft X-ray excess, a component that is almost ubiquitously seen in the X-ray spectra of unobscured, radiatively efficient active galaxies.
Previous X-ray observations of RX J1301.9+2747 in 2000 and 2009 strongly suggest that QPEs have been present for at least the past 18.5 years.
The detection of QPEs from a second galactic nucleus after GSN 069 rules out contamination by a Galactic source in both cases, such that QPEs ought to be considered a novel extragalactic phenomenon associated with accreting supermassive black holes.
   }

\keywords{galaxies: active --- galaxies: nuclei --- quasars: general --- quasars: super massive black holes --- X-rays: individuals: RX J1301.9+2747}

\titlerunning{X-ray QPEs in RX J1301.9+2747}
\authorrunning{Giustini, Miniutti \& Saxton}

\maketitle
\section{Introduction}\label{sec:intro}

The phenomena known as X-ray quasi-periodic eruptions (QPEs) have recently been detected  in the nucleus of the galaxy GSN 069 \citepads{2019Natur.573..381M} as high-amplitude, recurrent X-ray flares over a generally stable flux level (quiescent level).
In GSN 069, QPEs last about one hour and recur about every nine hours, with a subtle alternating pattern of long-short recurrence times and strong-weak amplitudes.
The QPE amplitude in GSN 069 is energy-dependent and up to two orders of magnitude in the $600-800$ eV band.
The X-ray spectrum of GSN 069 in the quiescent level is super-soft (most of the emission is at $E< 2$ keV) and it can be described via the thermal emission of an accretion disk with $kT\sim 50$ eV. During the overall X-ray decay between December 2010 and January 2019, the long-term evolution of the quiescent emission is consistent with the $L \propto T^4$ relation expected from a constant-area emitting accretion disk. This allowed \citetads{2019Natur.573..381M}  to estimate a black hole mass of $M_{BH}\sim 4\times 10^5 M_{\odot}$, associated with an uncertainty factor of a few due to the unknown black hole spin and observer inclination.
During QPEs, the X-ray spectrum of GSN 069 smoothly evolves into a warmer state with $kT\sim 120$ eV and back to the temperature preceding the QPE onset.
The QPEs detected in GSN 069 are a new phenomenon
whose physical origin is under investigation: they might be related, among other possibilities, to radiation- or magnetic-pressure instabilities of the inner accretion flow or to the orbital motion of a secondary body \citep{2019Natur.573..381M}.

Motivated by this discovery, we scanned the literature for cosmic sources that could be potentially analogous to GSN 069 in order to search for similar events.
The properties that make GSN 069 stand out among the general active galactic nuclei (AGN) population are: (i) small black hole mass; (ii) high Eddington ratio; (iii) pure thermal disk spectrum with little or no hard X-ray power law emission; and (iv) lack of broad optical or UV emission lines  \citep{2013MNRAS.433.1764M,2019Natur.573..381M}.
We selected RX J1301.9+2747 as a promising cosmic analogue candidate of GSN 069 on the basis of remarkably similar observational properties.
RX J1301.9+2747 is an edge-on post-starburst galaxy at $z = 0.02358$, which is possibly a member of a small group of four galaxies that lies $\sim 7'$ away from the center of the Coma cluster.
It had already been detected by EXOSAT in the 80s \citep[$f_{0.02-2.5\,\rm{keV}}\sim 1.4\times 10^{-12}$ erg cm$^{2}$ s$^{-1}$,][]{1985MNRAS.216.1043B} but it was only after the ROSAT observations performed in the 90s that the source was identified as an active galaxy \citep{2000MNRAS.318..309D}.

 A rapid flare lasting $\gtrsim 2 $ ks, with a variation of the X-ray count rate of a factor of $\sim 2.5$ with respect to the average value, was detected in the ROSAT light curve of RX J1301.9+2747 in June 1991\footnote{The actual amplitude and duration of the event are unknown, as only the decaying phase was caught by the satellite due to orbital constraints.} \citep[see Figure 4 of][]{2000MNRAS.318..309D}.
RX J1301.9+2747 was observed again by XMM-Newton in December 2000 and by Chandra in June 2009 and both observations confirmed its interesting timing properties. In particular, the 2000 EPIC-MOS observations caught one and a half events that are strikingly similar to the QPEs detected in GSN 069. In addition, a similar single X-ray flare was detected during the short ($\sim 5$ ks) Chandra observation nine years later \citep[see Figure 3 of][]{2013ApJ...768..167S}.
RX J1301.9+2747 is an ultra-soft X-ray source: if modeled with a blackbody, the ROSAT spectrum gives a temperature $kT\sim 55$ eV \citep{2000MNRAS.318..309D}.
The flux-resolved spectral analysis performed on the XMM-Newton and Chandra data by \citet{2013ApJ...768..167S} and \citet{2017ApJ...837....3S} revealed a spectrum that is well-fitted by a thermal disk component with $kT\sim 30-50$ eV in the low-flux state, and $kT\sim 100-300$ eV in the high-flux state, plus a weak hard power law emission.
The Eddington ratio of RX J1301.9+2747 is estimated to be at the level of  $\dot{m}\sim 0.14$, and its black hole mass at  $M_{BH}\sim 0.8-2.8 \times 10^6 M_{\odot}$, from the UV/X-ray analysis performed by \citet{2017ApJ...837....3S}, which also revealed an absence of broad optical/UV emission lines.
\citetads{2015MNRAS.446.1312M} noted the peculiar behaviour of the XMM-Newton 2000 X-ray light curve of RX J1301.9+2747 and proposed an explanation of the narrow flare based on a disk emission leaking through small "windows" in a rotating, optically thick structure, such as, a type of wind.
By performing a phase-resolved  analysis, \citetads{2015MNRAS.446.1312M} found a smooth evolution of spectral properties during the flare, similar to that observed in GSN 069 by \citetads{2019Natur.573..381M}

Based on these observational properties that so strikingly resemble those of GSN 069, we asked for a Director's discretionary time (DDT) XMM-Newton observation of RX J1301.9+2747 that was performed on 30 and 31 May 2019, whose scientific results  are reported in this Letter.
We present the data reduction and analysis results in Section \ref{sec:data}, the discussion in Section \ref{sec:discu}, and we present our conclusions in Section \ref{sec:conclu}.
Errors are quoted at the $1\sigma$ confidence level throughout the paper.
A flat cosmology ($\Lambda=0.73$, $q_0=0$, $H_0=70$ km s$^{-1}$ Mpc$^{-1}$) is assumed for the computation of the source intrinsic luminosity.


\section{Data reduction and analysis}\label{sec:data}

\begin{table*}
\caption{Observation log for the XMM-Newton EPIC-pn observations of RX J1301.9+2747 performed in 2019 (full frame mode, thin optical filter) and 2000 (full frame mode, medium optical filter). Dates of observations are in Coordinated Universal Time (UTC).\label{table:obslog}}
\centering
\begin{tiny}
\begin{tabular}{ccccccc}
\hline \hline
OBSID & Start/End date &  exposure & (src+bkg)$_{0.2-2}$ & (bkg)$_{0.2-2}$ & (src+bkg)$_{2-10}$ & (bkg)$_{2-10}$ \\
 &      (yyyy-mm-dd hh:mm:ss) & (s)  &  counts $s^{-1}$ & counts $s^{-1}$ & counts $s^{-1}$ &counts $s^{-1}$ \\
\hline
0851180501 & 2019-05-30 20:42:24/2019-05-31 10:09:04 & 45170 & $0.171\pm{0.002}$ & $0.0063\pm{0.0005}$ & $0.0046\pm{0.0004}$ & $0.0044\pm{0.0004}$\\
0124710801  & 2000-12-10 20:13:20/2000-12-11 04:30:05 & 23750 & $0.119\pm{0.003}$ & $0.0051\pm{0.0005}$ & $0.0032\pm{0.0004}$ & $0.0038\pm{0.0005}$  \\
\hline
\end{tabular}
\end{tiny}
\end{table*}

We reduced and analyzed the new data collected by XMM-Newton in May 2019, taking place during a  DDT observation pointed at RX J1301.9+2747 (OBSID: 0851180501).
We also re-analyzed, using calibration files generated in June 2019, the XMM-Newton data collected in December 2000, during an observation pointed at the Coma Cluster, which included RX J1301.9+2747 in the field of view (OBSID: 0124710801, PI: F. Jansen).
For the data reduction and analysis, we used the XMM-Newton Science Analysis System (SAS) v.17.0.0, following standard SAS threads as recommended by the XMM-Newton Science Operation Centre, and the HEASoft v.6.22.1 with Xspec v.12.9.1p.
For the 2019 observation, the whole exposure is retained for the light-curve analysis, while the last $6.5$ ks of the exposure are discarded for the spectral analysis, due to large background flares. The whole 2000 dataset is used for the scientific analysis.
In both epochs of observation the signal-to-noise ratio (S/N) drops above 2 keV, therefore, for our analysis we discard the signal above this energy.
We focus our analysis on the EPIC-pn data because of superior S/N with respect to the EPIC-MOS cameras.
Details of the observations are reported in Table~\ref{table:obslog}.


\subsection{Light curve analysis}

\begin{figure*}[ht!]
\centering
\includegraphics[ width=16cm]{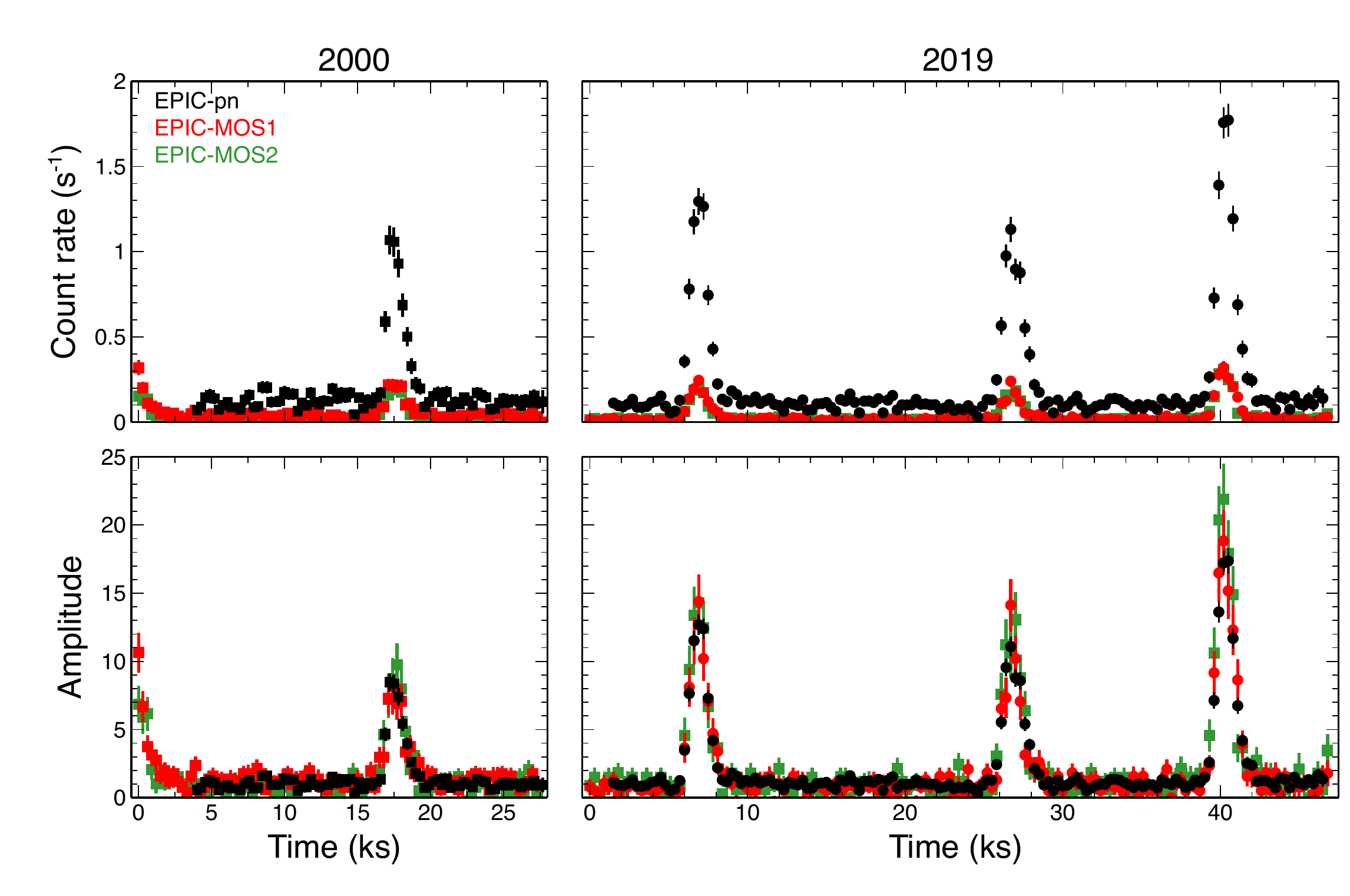}
\caption{Background-corrected light curves of RX J1301.9+2747, extracted with time bins of 300 s in the $0.2-2$ keV band during the 10-11 December 2000 (left panels) and the 30-31 May 2019 (right panels) XMM-Newton observations. Bottom panels: the count rates have been normalised to the quiescent level.
In black: EPIC-pn, in red: EPIC-MOS1, in green: EPIC-MOS2 data.\label{fig:2019lcurve}}
\end{figure*}

The RX J1301.9+2747 light curves extracted  in the $0.2-2$  keV band are shown in Fig. \ref{fig:2019lcurve}.
During the 2019 observation, three rapid flares were detected by all three EPIC cameras $\sim 7$, $\sim 27$, and $\sim 40$ ks after the beginning of the scientific exposure, with the count rate sharply increasing and decreasing with respect to a stable X-ray emission.
During the 2000 observation, one and a half events were detected by the EPIC-MOS cameras, while only one was detected by the EPIC-pn due to the later start of its scientific exposure.
In both epochs of observation, the shape of the X-ray light curve of RX J1301.9+2747 was very different from the typical AGN, in addition to being remarkably similar to the one displayed by GSN 069 from December 2018 onward \citep{2019Natur.573..381M}.
Although the events in RX J1301.9+2747 do not seem to recur quasi-periodically, we call these events QPEs, using the same nomenclature introduced by \citetads{2019Natur.573..381M} for GSN 069. This choice is based on the strikingly analogy between the X-ray light curves of RX J1301.9+2747 and GSN 069 (including the remarkably stable quiescent level), along with the body of similarities between the two sources (see Section 1) and the further almost identical spectral and timing behaviour (see below) which, altogether strongly suggest that the phenomena observed in RX J1301.9+2747 and GSN 069 are, in fact, one and the same.

\begin{figure*}
     \centering
        \includegraphics[width=16.cm]{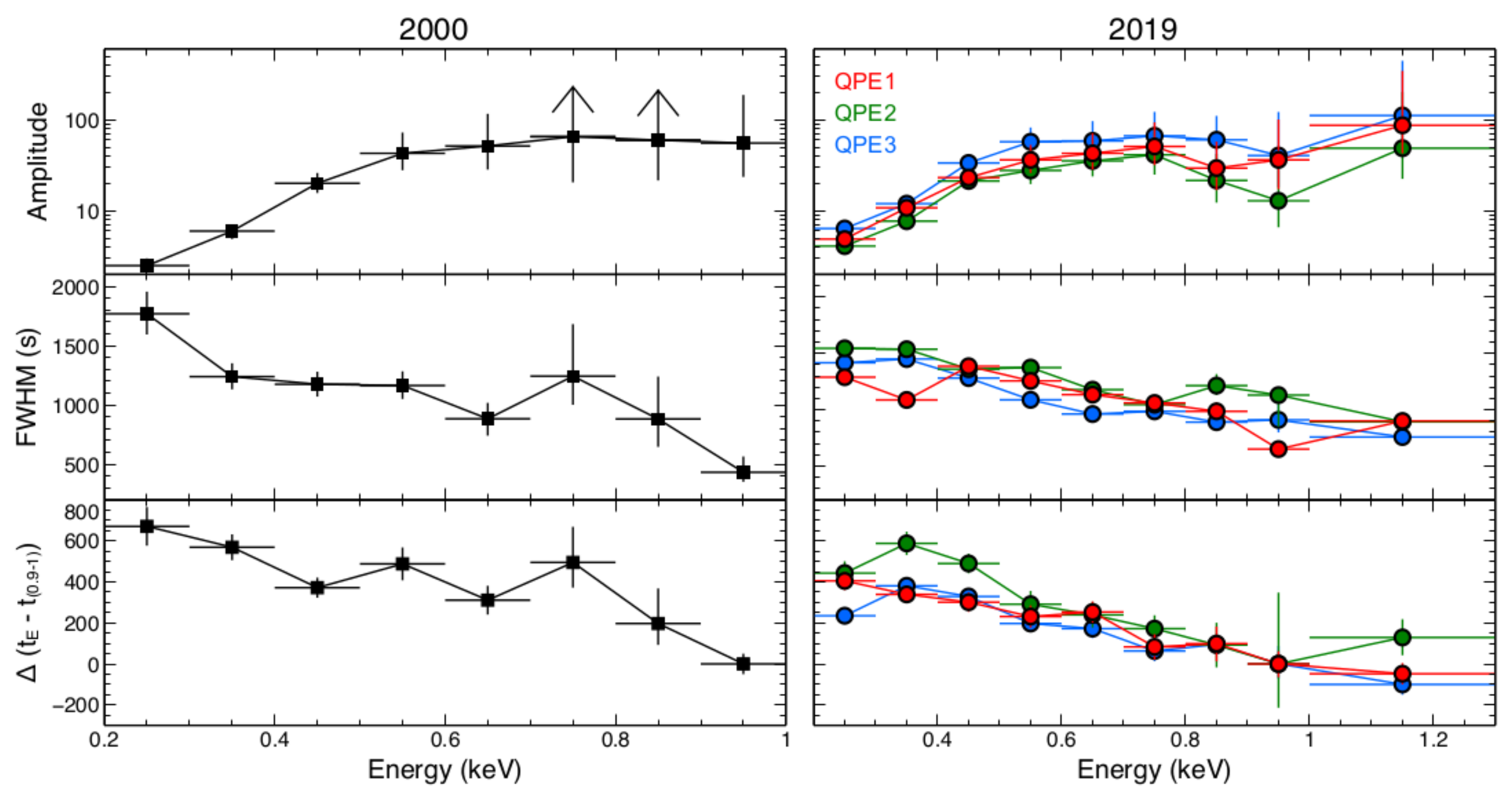}
     \caption{RX J1301.9+2747 QPE properties: amplitude (Gaussian intensity over quiescent count rate), duration (Gaussian FWHM), and peak time (Gaussian centroid) with respect to that measured in the $0.9-1$ keV band. The 2000 QPE is plotted with black squares, the 2019 QPEs are plotted with circles (QPE1 in red, QPE2 in green, and QPE3 in blue).}
     \label{fig:QPE}
 \end{figure*}

In order to explore the energy-dependent properties of the QPEs of RX J1301.9+2747, we model its light curves with a constant representing the quiescent flux level, plus Gaussian emission lines representing the QPEs.
We then compute the QPE amplitude as the ratio of the the Gaussian normalization and the constant count rate; the QPE duration as the full width at half maximum (FWHM) of the Gaussian line; and the QPE peak position as the Gaussian centroid.
We measure these properties in the light curves extracted in nine contiguous energy bands, from 0.2 to 1.3 keV (Fig. \ref{fig:LCURVES}); for the 2000 QPE analysis, we discard the last energy band, as at $E > 1$ keV the S/N drops.
For completeness, we also plot in the bottom panels of Fig. \ref{fig:LCURVES} the 1.3-2 keV light curves for the two epochs of observation.
Results of the analysis are shown in Fig. \ref{fig:QPE}, where the 2000 QPE properties are plotted with black squares, the three 2019 QPEs (hereafter QPE1, QPE2, and QPE3) with red, green, and blue circles, respectively.
In both 2000 and 2019, the X-ray QPEs in RX J1301.9+2747 are more intense when measured at higher energies: their amplitude is $\ll 10$ at $E\lesssim 0.3$ keV , but quickly reaches a factor $\gtrsim 50$ at $E\gtrsim 0.5$ keV, which remains more or less constant up to the highest energies probed.
The S/N drops at $E>1 $ keV in 2000, while in 2019 there is good S/N up to $E\sim 1.3$ keV, where the QPEs have the largest amplitude.
The duration of the QPEs is very short: on average $\sim 1200$ s (FWHM), and as short as $500-800$ s at the highest energies probed. Indeed, the duration of the QPEs in RX J1301.9+2747 increases when measured at lower energies, both in 2000 and 2019.
Also, the arrival time of the QPE increases when measured at lower energies, both in 2000 and in 2019: the difference in Gaussian peaks used to model the QPE between the $0.9-1$ keV energy band is $> 500$ s at $E\lesssim 0.5$ keV.

The three QPEs detected in 2019 have different amplitudes, with QPE3 being more intense than QPE 1, which, in turn, is more intense than QPE2 at all energies.
QPE2 is the one with the lowest amplitude and also the one with the longest duration at all energies.

 \subsection{Spectral analysis}

\begin{figure*}[ht!]
\centering
\includegraphics[ width=17.5cm]{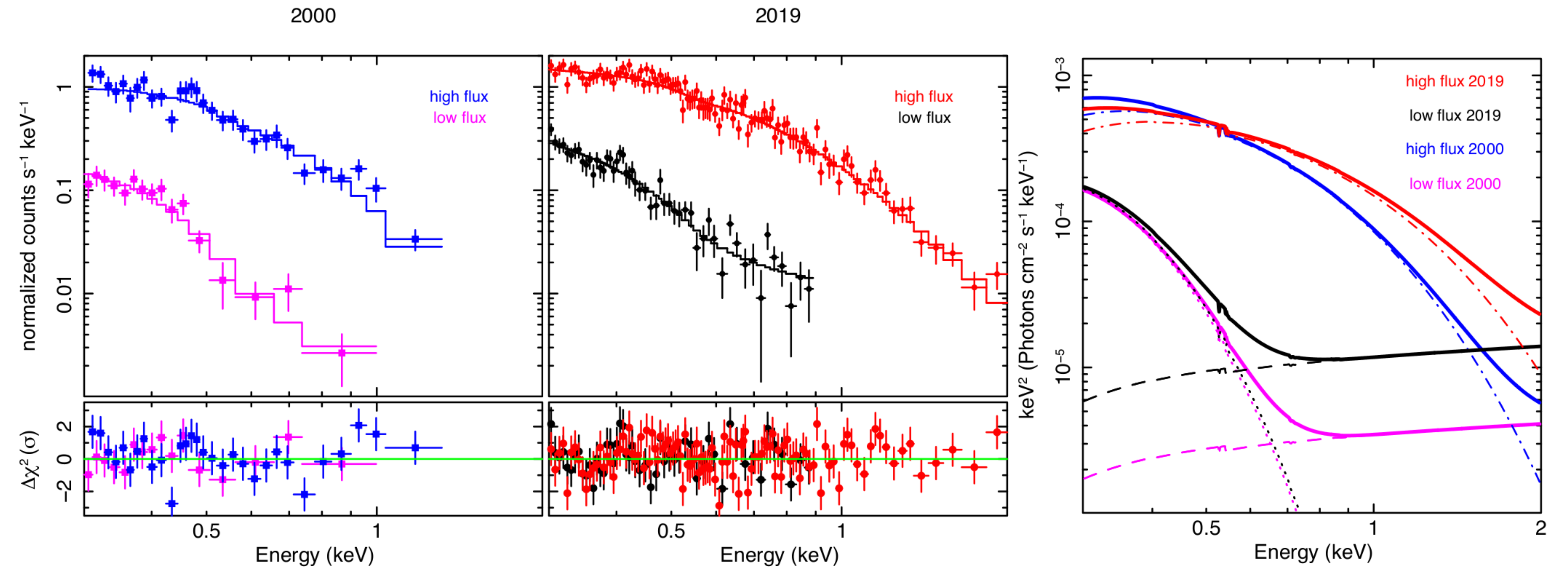}
\caption{Left and middle panels: flux-resolved EPIC-pn spectra of RX J1301.9+2747 in the 2000 and 2019 observations, along with spectral residuals to model 3. Right panel: model 3; the dashed line is the power law, the dotted line the disk blackbody (both kept tied within the low-flux and high-flux state of each observation), and the dashed-dotted line the bremsstrahlung emission emerging in the high-flux spectra.\label{fig:eeuf}}
\end{figure*}

 We divide the  data in a low-flux and a high-flux state, using a threshold of $<0.2$ ct s$^{-1}$ and $>0.4$ ct s$^{-1}$  applied to the $0.2-2$ keV EPIC-pn light curves of both the 2000 and the 2019 observations.
In the low-flux state, the signal is dominated by the background at $E  \gtrsim 0.9$ keV, therefore, the data above those energies are discarded; in the high-flux state the spectra are harder and the source is instead well-detected up to $E\sim 2$ keV.
At low energies, data below $0.3$ keV are discarded due to remaining calibration uncertainties.
The background-subtracted spectra are grouped with a  minimum of 20 counts per bin, and the $\chi^2$ statistics is used as a measure of the goodness of fit of the model to the data.
Spectral analysis results are reported in Table~\ref{table:spectralfit}.

In all our fits, we model the absorption along the line of sight with the  $\texttt{tbabs}$ model in Xspec \citepads{2000ApJ...542..914W}, leaving the column density value $N_H$  free to vary.
First, we fit the data to a simple thermal disk model (model 0: \texttt{diskbb} into Xspec), leaving all the parameters free to vary during the fit: the resulting statistics are very poor, giving
 a reduced chi squared $\chi^2_r \sim 1.6$ for 202 degrees of freedom ($\nu$). The addition of a power law component improves  the fit statistics significantly ($\Delta\chi^2/\Delta\nu=94/4$). The power law slope is found to be unconstrained, therefore, in all epochs and in all the subsequent models, we fixed its value to $\Gamma=1.8$ \citepads[a value that is typical for the intrinsic continuum in AGN, see, e.g.,][]{2005A&A...432...15P}.
The fit to this model (model 1: \texttt{powerlaw + diskbb} in Table \ref{table:spectralfit}) is marginally acceptable ($\chi^2_r \sim 1.15$) for a disk temperature of about $60$ eV in the low-flux state at both epochs, and $\sim 100-150$ eV in the high-flux state.

We then assume a scenario where the quiescent emission stays constant during each observation and a further emission component is only added to the high-flux state to represent the spectral contribution of the QPE.
In practice, the model parameters for the quiescent emission (modeled with \texttt{powerlaw + diskbb}) are tied between the low-flux and high-flux spectra of each epoch, while they are allowed to vary between 2000 and 2019; the model parameters of the spectral component representing the QPE emission are fixed to zero in the low-flux spectra, and are left free to vary in the high-flux spectra.
We use three different models to represent the emerging QPE during the high-flux state: a blackbody emission (Model 2: \texttt{bbody}), a bremsstrahlung emission (model 3: \texttt{bremss}), and a Comptonization emission where the disk photon temperature is used as input for the Comptonizing region. We use two different models for the Comptonized disk emission: \texttt{nthcomp} (model 4a) by \citetads{1996MNRAS.283..193Z} and \citetads{1999MNRAS.309..561Z}, and \texttt{comptt} (model 4b) by \citetads{1994ApJ...434..570T}.
Model 2 does not give a fair representation of the data ($\chi^2_r\sim 1.4$): strong residuals are evident across all the energy ranges analyzed and especially at the highest energies, where the model severely underestimates the data.
Both Models 3 and 4 give, instead, a fair fit to the data, with $\chi^2_r < 1.1$. For simplicity, we refer to Model 3 as our best-fit model for presenting figures, although we note that the spectral shape of the component emerging during the high-flux state is equivalently well-represented by Model 4a and 4b.
The only significant differences are the unabsorbed luminosities of the disk and QPE spectral components, which are larger by about, respectively, 60-30\% in Models 4 than in Model 3 because the Comptonization models allow for a slightly larger neutral absorbing column density (Table \ref{table:spectralfit}); in the following we quote Model 3 luminosities.

In Figure \ref{fig:eeuf} , we show the 2000 (left panel) and 2019 (middle panel) spectra of RX J1301.9+2747 in the low-flux and high-flux state fitted to model 3, along with spectral residuals. In the right panel of Fig. \ref{fig:eeuf} we plot the corresponding four theoretical models.
In both Models 3 and 4, the quiescent spectra of the low-flux state are well-described by a disk emission with temperature $kT^{disk}\sim 50$ eV and a constant flux between epochs, $f^{disk}_{0.3-2}\sim 8\times 10^{-14}$ erg cm$^{-2}$ s$^{-1}$, corresponding to an unabsorbed luminosity $L^{disk}_{0.3-2}\sim 1.5\times 10^{41}$ erg s$^{-1}$. The flux of the hard X-ray power law in quiescence is instead found to be significantly higher in 2019 ($f^{pow}_{0.3-2}\sim 3.3\times 10^{-14}$ erg cm$^{-2}$ s$^{-1}$ , corresponding to $L^{pow}_{0.3-2}\sim 4.5\times 10^{40}$ erg s$^{-1}$) than in 2000 ($f^{pow}_{0.3-2}\sim 10^{-14}$ erg cm$^{-2}$ s$^{-1}$, or $L^{pow}_{0.3-2}\sim 1.3\times 10^{40}$ erg s$^{-1}$).
The QPE emergent spectrum (i.e., the difference between the high- and low-flux state) is well-represented by either a thermal bremsstrahlung or by a Comptonized emission with an observed flux $f^{QPE}_{0.3-2}\sim 7.5\times 10^{-13}$ erg cm$^{-2}$ s$^{-1}$ and a corresponding luminosity (corrected for absorption along the line of sight) of $L^{QPE}_{0.3-2}\sim 1.2\times 10^{42}$ erg s$^{-1}$.
In model 3, the bremsstrahlung temperature is found to increase between 2000 and 2019, from $\sim 220$ to $\sim 300$ eV.
In model 4a, it is the Comptonisation asymptotic power law slope to significantly change, hardening from $\Gamma^{nth}\sim 4.4$ to $\sim 3$ from 2000 to 2019.
In model 4b, the Comptonization region optical depth is found to increase from $\sim 7$ to $\sim 14$ from 2000 to 2019.
In Model 3, a marginal improvement of the fit ($\Delta\chi^2=5$ for two extra degrees of freedom) is obtained by allowing also the normalization of the power law to vary in the high-flux state.
In this case, the temperature of the bremsstrahlung component adjusts to lower temperatures, but nonetheless there is still a significant increase in value between 2000 ($kT^{bre}=173^{+23}_{-19}$ eV) to 2019 ($kT^{bre}=277^{+23}_{-20}$ eV).
No improvement of the fit to models 4a and 4b is found instead when allowing the power law normalization to vary during the high-flux state, because the power law normalization is degenerate with the slope of the asymptotic power law of the \texttt{nthcomp} component (model 4a) and with the optical depth of the \texttt{comptt} component (model 4b).
We point out that no spectral model can explain the low-flux and high-flux spectra with only a change in overall normalization because of the different spectral shape at the two flux levels.
We conclude that the 2019 QPE spectrum is harder and hotter than the 2000 QPE spectrum, regardless of the adopted best-fitting model.

The amount of intrinsic neutral absorption at the redshift of RX J1301.9+2747 is found to be comparable to the one measured by the Leiden/Argentine/Bonn survey \citepads[$N_H = 8.2 \times 10^{19}$ cm$^{-2}$,][]{2005A&A...440..775K} in model 3, while it is significantly larger in Model 4, which allows for a softer intrinsic spectral shape.
With the present data quality, we cannot exclude the presence of more substantial columns of gas along the line of sight, especially if this is either ionized or only partially covering the continuum source.

\section{Discussion}\label{sec:discu}

RX J1301.9+2747 is only the second extragalactic source where rapid, intense, and repeated X-ray flares have been detected, after GSN 069 \citepads{2019Natur.573..381M}.
Given the striking similarities between the X-ray properties of the two sources which strongly suggest that we are dealing with the same phenomenon, we refer to the X-ray flares detected in RX J1301.9+2747 as quasi-periodic eruptions, or QPEs (Fig. \ref{fig:2019lcurve}).

 As in GSN 069, the QPEs in RX J1301.9+2747 are very short and very intense and they are shorter, more intense, and peak earlier when measured at higher energies (Figs. \ref{fig:LCURVES} and \ref{fig:QPE}).
 However, at the highest energies probed, the X-ray QPEs in GSN 069 are roughly twice as long as the QPEs in RX J1301.9+2747, which are as short as ten to fifteen minutes in FWHM.
An upper limit on the size $S$ of the X-ray emitting region is given by the distance that light can travel in a given time $\Delta t$, and yields:
 $S < 200\,(\Delta t / 1000\,$s$)\,(10^6\,M_{\odot}/M_{BH})\:r_g$, where $r_g=GM_{BH}/c^2$ is the gravitational radius.
 Since, at the highest probed energies, the count rate doubling time during QPEs in RX J1301.9+2747 is of the order of 300 seconds at most, we are probing an X-ray emitting region of the order of few tens of gravitational radii.

 The amplitude of the QPEs in RX J1301.9+2747 is as low as $2-6$ at $E\lesssim 300$ eV (Fig. \ref{fig:QPE}) and, as in GSN 069, it drops at lower energies: in RX J1301.9+2747, there is no variability detected in the light curve extracted with the UVW2 filter ($E\sim 5.8$ eV) of the optical monitor onboard XMM-Newton, and no variability is observed with the UVM2 filter ($E\sim 5.4$ eV) in GSN 069. Therefore, if related to accretion, QPEs must originate in the X-ray-emitting-only portion of the accretion flow, very close to the central SMBH in RX J1301.9+2747.
 The dynamical timescale at $10 r_g$ around a $10^6M_{\odot}$ black hole is $t_{dyn}\sim 160$ s.
 The three QPEs detected in the 2019 observation of RX J1301.9+2747 define two recurrence times: one long of $\sim 20$ ks between QPE1 and QPE2, and one short of $\sim 13.5$ ks between QPE2 and QPE3.
 The QPE amplitude also varies, with QPE1 and QPE3 being stronger than QPE2.
The recurrence times between the QPEs can be compared to the viscous timescale in an accretion flow, $t_{vis}=(H/R)^{-2}\,t_{th}$.
Here $H/R$ is the scale height of the accretion flow and $t_{th}=t_{dyn}/\alpha$ is the thermal timescale, that can be associated to the rising and decaying times of the QPEs.
Given the short rising and decaying times in RX J1301.9+2747 of about 1 ks, a large viscosity parameter $\alpha\sim 0.15$ would be inferred.
Adopting this value for the viscosity parameter yields $H/R\sim 0.25$, possibly indicating a geometrically thick or magnetically elevated inner accretion flow \citepads{2018MNRAS.480.3898N,2019MNRAS.483L..17D}.

The quiescent X-ray spectra of both RX J1301.9+2747 and GSN 069 are super-soft and well-represented by thermal accretion disk emission with a temperature of about 50 eV and a $0.2-2$ keV luminosity of about $10^{41}$ erg s$^{-1}$.
The typical AGN X-ray signatures are very weak (e.g., the hard X-ray power law) or completely absent (e.g., a reflection component, a soft X-ray excess) in both sources in the quiescent level. In GSN 069, the hard X-ray power law emits a very low luminosity ($\lesssim 10^{40}$ erg s$^{-1}$) compared to typical AGN and has a roughly constant amplitude in different epochs of observation. A similar luminosity level is inferred for the hard X-ray power law of RX J1301.9+2747 during the 2000 XMM-Newton observation, while in 2019, its luminosity increased significantly, by a factor of $\sim 3$.
The RX J1301.9+2747 quiescent spectra are instead remarkably stable in flux over 18.5 years; there is however a surplus of photons with energy $E\gtrsim 0.5$ keV in 2019 compared to 2000 (see Fig. \ref{fig:eeuf}).

During QPEs, a component with an unabsorbed luminosity $L_{0.3-2}\sim 1.2-1.6\times 10^{42}$ erg s$^{-1}$ emerges in the spectra of RX J1301.9+2747. This luminosity is comparable to the one of the QPEs in GSN 069.
While in GSN 069, the QPE spectral shape is well-represented by a blackbody emission (even though Comptonization or bremsstrahlung give comparable fits; Miniutti et al., in preparation), in RX J1301.9+2747, its spectral shape looks harder than a simple blackbody and can be well-represented instead by a thermal bremsstrahlung with a temperature of $200-300$ eV, or by Comptonization of the seed disk photons into a warm gas with a similar temperature of the bremsstrahlung model.
From 2000 to 2019, the RX J1301.9+2747 QPE spectrum has also changed, implying either an increase of the bremsstrahlung emitting gas temperature or a hardening of the Comptonized emission modeled with \texttt{nthcomp}. This hardening would correspond, for a fixed temperature, to an increase of the Comptonizing region optical depth from $\tau \sim 5$ to $\tau\sim 15$ from 2000 to 2019 \citepads{2018A&A...611A..59P}, compatible with the results of our spectral analysis with \texttt{comptt}. The Comptonizing region in RX J1301.9+2747 would be optically thick at all epochs of observation.

When measured at $E\lesssim 0.4$ keV, the amplitude of the 2000 QPE is smaller than the amplitude of the 2019 QPEs, while it is slightly larger when measured at $E\sim 0.8-1$ keV. This is because the underlying quiescent spectrum is softer in 2000 than in 2019, therefore providing a larger number of photons below 400 eV, and a smaller number of photons above 800 eV, against which the QPE amplitude can be measured.
Above 1 keV, the S/N drops in 2000, while in 2019 there is good S/N until $E\sim 1.3 $ keV, and at these energies the QPE amplitude is the largest for all the three 2019 QPEs.

There is also a hint of anti-correlation between the intensity and the duration of the 2019 QPEs.
In particular, QPE2 has the lowest amplitude and the longest duration, while the opposite is true for QPE3, at all the energies probed.
Further, longer observations might allow the detection of more QPEs and improve the S/N sufficiently to assess
whether the differences in properties between QPEs are significant and whether there exists a correlation between the QPE amplitude and duration. Such future observations should also clarify the pattern of variability of the QPEs of
RX J1301.9+2747, and allow a more thorough comparison with those of GSN 069.

\section{Conclusions}\label{sec:conclu}

During a 48 ks XMM-Newton observation performed in May 2019, three strong and rapid X-ray QPEs have been detected in the nucleus of the galaxy RX J1301.9+2747.
These QPEs seem to be long-lived:
in fact, about 1.5 QPEs with similar properties to those of 2019 were detected in a 2000 archival XMM-Newton observation of RX J1301.9+2747 \citepads{2013ApJ...768..167S}, and also observations performed by ROSAT in 1994 \citepads{2000MNRAS.318..309D} and by Chandra in 2009 \citepads{2017ApJ...837....3S} revealed interesting X-ray variability properties, with sudden increases or decreases in X-ray count rate above a stable low-flux level.

The general properties of the X-ray QPEs observed in RX J1301.9+2747 are similar to those of the QPEs observed in the discovery source GSN 069 \citepads{2019Natur.573..381M}: their merged spectrum looks like a thermal component (with a temperature of about 100-300 eV, depending on spectral modeling), with a $0.2-2$ keV intrinsic luminosity of the order of $10^{42}$ erg s$^{-1}$, about one order of magnitude higher than the luminosity of the quiescent level.
There are also clear differences between the QPEs observed in RX J1301.9+2747 and GSN 069.
Not only the QPEs in RX J1301.9+2747 are shorter than those in GSN 069, but their time separation looks generally also shorter.
The three events observed in RX J1301.9+2747 during the 2019 observation do not define a quasi-period, as the recurrence times are significantly different ($\sim 20$ and $\sim 13.5$ ks). We point out that the same also occurs in GSN 069, only with a much smaller difference between long and short recurrence times\footnote{The maximum difference between consecutive recurrence times observed so far in GSN 069 is $\sim 3.4$ ks (Miniutti et al., in preparation).}.
On the other hand, if the two sources do share the same physical phenomenon, as strongly suggested by our analysis, longer observations of RX J1301.9+2747 will reveal whether the
quasi-periodic behaviour has similar odd-even QPE pairs to those of GSN 069.

 The quiescent spectrum of RX J1301.9+2747 is well-described by a thermal disk with a temperature of 50 eV and a $0.2-2$ keV luminosity of about $10^{41}$ erg s$^{-1}$, constant between the 18.5 years between the two XMM-Newton observations; plus a weak hard X-ray power law, whose $0.2-2$ keV luminosity more than tripled between 2000 and 2019, when it is still, however, $< 5\times 10^{40}$ erg s$^{-1}$.
Also the spectrum of the QPE changed between 2000 and 2019, having become harder: this might mean that in the time elapsed between the two observations the temperature of the QPE has increased or that in 2019, the power law emission was also contributing to the QPE, which is contrary to 2000.

While in GSN 069, the QPEs are detectable during the overall $\sim 10$ year-long (so far) decay following an outburst first detected in 2010, the QPEs of RX J1301.9+2747 are detectable during two observations $\sim 18.5$ years apart and at a similar flux or luminosity level. Very long-lived tidal disruption events may perhaps explain the long-term evolution of both sources \citepads[e.g.,][]{2012ApJ...757..134M,2017NatAs...1E..33L}.
From a phenomenological point of view, the QPE spectral evolution can be described as a transient and fast transition from a disk-dominated to a soft excess-dominated state and, if so, QPEs may provide crucial clues on the origin of this X-ray spectral component which is almost ubiquitous in unobscured, radiatively efficient AGN.
The question whether QPEs are directly associated with accretion flow variability or instabilities or whether they are due instead to extrinsic phenomena (such as interactions with a secondary orbiting body) remains to be studied \citepads[see][for possible interpretations]{2020MNRAS.493L.120K,2020arXiv200207318C}.
Future X-ray observations of both sources will enable us to constrain possible theoretical models taking advantage of the different properties and timescales in the two sources, which need to be consistent with a similar theoretical framework.

 The detection of X-ray QPEs in RX J1301.9+2747 doubles the number of galactic nuclei where this new phenomenon has been observed after their discovery in GSN 069.
This rules out contamination by a Galactic source in both cases, assessing QPEs as a novel extragalactic phenomenon associated with supermassive accreting black holes.

\begin{acknowledgements}
We thank the XMM-Newton Project Scientist N. Schartel for approving the DDT observation, and the XMM-Newton staff for performing it.
We also thank the referee for useful comments and suggestions.
MG is supported by the ``Programa de Atracci\'on de Talento'' of the Comunidad de Madrid, grant number 2018-T1/TIC-11733.
GM is supported by the Spanish State Research Agency (AEI) Project No. ESP-2017-86582-C4-1-R.
This research has been partially funded by the AEI Project No. MDM-2017-0737 Unidad de Excelencia ``Mar\'ia de Maeztu'' - Centro de Astrobiolog\'ia (INTA-CSIC).
\end{acknowledgements}

\bibliographystyle{aa}

\begin{thebibliography}{}


\bibitem[Branduardi-Raymont et al. (1985)]{1985MNRAS.216.1043B}
{{Branduardi-Raymont}, G., {Mason}, K.~O., {Murdin}, P.~G., \&
  {Martin}, C.} 1985,
MNRAS, 216, 1043

\bibitem[Coughlin \& Nixon (2020)]{2020arXiv200207318C}
{Coughlin, E. R. \& Nixon, C. J.} 2020, arXiv e-prints, arXiv:2002.07318

\bibitem[Dewangan et al. (2000)]{2000MNRAS.318..309D}
{{Dewangan}, G.~C., {Singh}, K.~P., {Mayya}, Y.~D., \&
  {Anupama}, G.~C.} 2000, MNRAS, 318, 309

  \bibitem[Dexter \& Begelman (2019)]{2019MNRAS.483L..17D}
  {Dexter, J. \& Begelman, M. C. } 2019, MNRAS, 483, L17

\bibitem[Kalberla et al. (2005)]{2005A&A...440..775K}
       {{Kalberla}, P.~M.~W., {Burton}, W.~B., {Hartmann}, et al.} 2005,
       A\&A, 440, 775

\bibitem[King (2020)]{2020MNRAS.493L.120K}
       {King, A.  } 2020, MNRAS, 493, L120

\bibitem[Lin et al. (2017)]{2017NatAs...1E..33L}
                     {{Lin}, Dacheng and {Guillochon}, James and {Komossa}, S., et al.} 2017,
                     Nature Astronomy, 1, 33

\bibitem[MacLeod et al. (2012)]{2012ApJ...757..134M}
  {{MacLeod}, M., {Guillochon}, J., \& {Ramirez-Ruiz}, E.} 2012,
        ApJ, 757, 134

        \bibitem[Middleton \& Ingram (2015)]{2015MNRAS.446.1312M}
          {{Middleton}, M. J. \& {Ingram}, A. R.} 2015,
                MNRAS,446, 1312


\bibitem[Miniutti et al. (2019)]{2019Natur.573..381M}
 {{Miniutti}, G., {Saxton}, R.~D., {Giustini}, M., et al.} 2019,
Nature, 573, 381


\bibitem[Miniutti et al. (2013)]{2013MNRAS.433.1764M}
 {{Miniutti}, G., {Saxton}, R.~D., {Rodr{\'\i}guez-Pascual}, P.~M., et al.} 2013,
MNRAS, 433, 1746

\bibitem[Noda \& Done (2018)]{2018MNRAS.480.3898N}
 {{Noda}, H. \& {Done}, C.,} 2018,
MNRAS, 480, 3898


\bibitem[Petrucci et al. (2018)]{2018A&A...611A..59P}
 {{Petrucci}, P. -O., {Ursini}, F., {De Rosa}, A., et al.} 2018,
A\&A, 611, 59

\bibitem[Piconcelli et al. (2005)]{2005A&A...432...15P}
 {Piconcelli, E., Jimenez-Bail\'on, E., Guainazzi, M., et al.}  2005,
A\&A, 432, 15

\bibitem[Shu et al. (2017)]{2017ApJ...837....3S}
 {{Shu}, X.~W., {Wang}, T.~G., {Jiang}, N., et al.} 2017,
ApJ, 837, 3

\bibitem[Sun et al. (2013)]{2013ApJ...768..167S}
 {{Sun}, L., {Shu}, X., \& {Wang}, T.} 2013,
ApJ, 768, 167

       \bibitem[Titarchuk (1994)]{1994ApJ...434..570T}
        {{Titarchuk}, L., } 1994,
       ApJ, 434, 570


       \bibitem[Wilms et al. (2000)]{2000ApJ...542..914W}
        {{Wilms}, J., {Allen}, A., \& {McCray}, R.,} 2000,
       ApJ, 542, 914

       \bibitem[Zdziarski et al. (1996)]{1996MNRAS.283..193Z}
        {{Zdziarski}, A.~A., {Johnson}, W.~N., \& {Magdziarz}, P.,} 1996,
       MNRAS, 283, 193

              \bibitem[{\.Z}ycki et al. (1999)]{1999MNRAS.309..561Z}
               {{{\.Z}ycki}, P. T., {Done}, C., \& {Smith}, D. A.,} 1999,
              MNRAS, 309, 561

 \end{thebibliography}

\begin{appendix}
 \section{Supplementary figures and tables}
\begin{figure*}
     \centering
     \includegraphics[height=15.7cm]{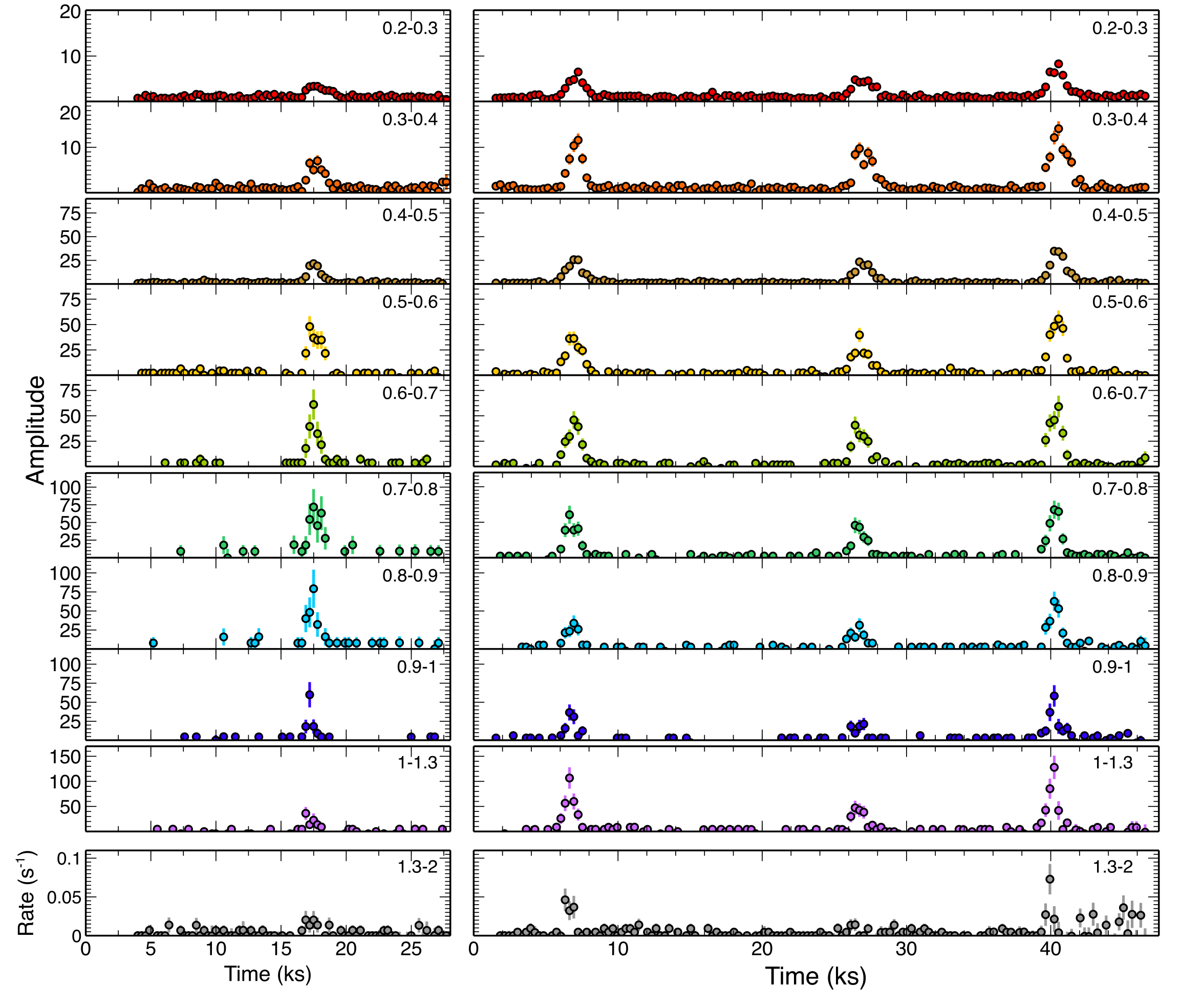}
     \caption{EPIC-pn light curves of RX J1301.9+2747 (December 2000 observation on the left, May 2019 observation on the right) binned to 300 s, corrected for the background and normalized to their quiescent count rate level, extracted in different energy bands labeled in units of keV in the top right corner of each panel. We note the different y-scales in different energy bands. In the bottom panels, we plot instead the light curve in the 1.3-2 keV band for the two epochs of observation.}
     \label{fig:LCURVES}
 \end{figure*}
\begin{table*}
\caption{RX J1301.9+2747: EPIC-pn spectral analysis results. The fluxes are observed erg cm$^{-2}$ s$^{-1}$. Errors are quoted at $1\sigma$ confidence level.}
\label{table:spectralfit}
\centering
\begin{tabular}{c c c c c  c c c }
\hline
\hline
\multicolumn{8}{c}{Model 1: \texttt{tbabs*(pow + diskbb)}: $\chi^2/\nu=227/198$} \\
\multicolumn{8}{c}{ $N_H< 3 \times 10^{19} $ cm$^{-2}$} \\
 Spectrum  & $f_{0.3-2}^{pow}$ &  $kT^{disk}$ & $f_{0.3-2}^{disk}$ & & & & \\
   & [erg cm s$^{-1}$] & [eV] &  [erg cm s$^{-1}$] & & & \\
2000 low & $8.7^{+3.9}_{-4.2}\times 10^{-15} $ & $58^{+1}_{-3} $ & $7.7^{+3.5}_{-0.8}\times 10^{-14}$  & & & & \\
2000 high  &  $1.9^{+0.3}_{-0.8}\times 10^{-13} $ & $103\pm{2}$  & $7.1^{+2.2}_{-1.0}\times 10^{-13}$ & & & & \\
2019 low & $3.1\pm{0.4}\times 10^{-14}$ &  $59^{+1}_{-3} $  & $8.0^{+2.9}_{-0.4}\times 10^{-14}$ & & & & \\
2019 high  & $1.2^{+0.1}_{-0.2}\times 10^{-13}$ & $149^{+4}_{-3}$  & $7.5^{+1.3}_{-0.2}\times 10^{-13} $ & & & & \\
\hline
\multicolumn{8}{c}{Model 2: \texttt{tbabs*(pow + diskbb + bbody)}: $\chi^2/\nu=295/200$}  \\
\multicolumn{8}{c}{ $N_H< 2 \times 10^{19} $ cm$^{-2}$}     \\
Spectrum  & $f_{0.3-2}^{pow}$ &  $kT^{disk}$ & $f_{0.3-2}^{disk}$ & $kT^{bb}$ &   $f_{0.3-2}^{bb}$ & &  \\
& [erg cm s$^{-1}$] & [eV] &  [erg cm s$^{-1}$] & [eV] & [erg cm s$^{-1}$] & \\
2000 low & \multirow{2}{*}{$1.0\pm{0.4}\times 10^{-14}$} & \multirow{2}{*}{$57\pm{1}$} & \multirow{2}{*}{$7.7^{+3.5}_{-1.2}\times 10^{-14}$} & $-$ & $-$ & & \\
2000 high & & & & $103\pm{4}$ & $6.9\pm{0.4}\times 10^{-13}$ & & \\
2019 low & \multirow{2}{*}{$3.9^{+0.3}_{-0.4}\times 10^{-14}$} & \multirow{2}{*}{$55^{+3}_{-1}$} & \multirow{2}{*}{$8.0^{+3.6}_{-0.5}\times 10^{-14}$} &  $-$ & $-$ & & \\
2019 high & & & & $125\pm{2}$ & $7.1^{+0.2}_{-0.1}\times 10^{-13} $ & & \\
\hline
 \multicolumn{8}{c}{Model 3: \texttt{tbabs*(pow + diskbb + brems)}: $\chi^2/\nu=217/200$}\\
\multicolumn{8}{c}{ $N_H  = 1.7\pm{0.9}\times 10^{20}$  cm$^{-2}$}     \\
Spectrum  & $f_{0.3-2}^{pow}$ &  $kT^{disk}$ & $f_{0.3-2}^{disk}$ & $kT^{bre}$ &   $f_{0.3-2}^{bre}$ & $f_{0.3-2}^{tot}$ &  \\
& [erg cm s$^{-1}$] & [eV] &  [erg cm s$^{-1}$] & [eV] & [erg cm s$^{-1}$] &  [erg cm s$^{-1}$] &\\
2000 low & \multirow{2}{*}{$9.5\pm{4.5}\times 10^{-15}$} & \multirow{2}{*}{$53^{+5}_{-4}$} & \multirow{2}{*}{$7.6^{+10.0}_{-4.2}\times 10^{-14}$} & $-$ & $-$ & $8.6^{+10.4}_{-5.3}\times 10^{-14}$ & \\
2000 high & & & & $222\pm{17}$ & $7.4^{+1.9}_{-1.4}\times 10^{-13}$ & $8.3^{+2.9}_{-1.9}\times 10^{-13}$  & \\
2019 low & \multirow{2}{*}{$3.2^{+0.6}_{-0.5}\times 10^{-14}$} & \multirow{2}{*}{$54\pm{4}$} & \multirow{2}{*}{$8.0^{+9.0}_{-4.1}\times 10^{-14}$} &  $-$ & $-$ & $1.1^{+1.0}_{-0.4}\times 10^{-13}$ & \\
2019 high & & & & $302\pm{17}$ & $7.7^{+1.4}_{-1.1}\times 10^{-13} $ & $8.8^{+2.4}_{-1.6}\times 10^{-13}$  & \\
\hline
 \multicolumn{8}{c}{Model 4a: \texttt{tbabs*(pow + diskbb + nthcomp)}: $\chi^2/\nu=209/198$}\\
\multicolumn{8}{c}{ $N_H  = 4\pm{1}\times 10^{20}$ cm$^{-2}$  cm$^{-2}$}     \\
Spectrum  & $f_{0.3-2}^{pow}$ &  $kT^{disk}$ & $f_{0.3-2}^{disk}$ & $kT^{nth}$ &  $\Gamma^{nth}$ & $f_{0.3-2}^{nth}$ & $f_{0.3-2}^{tot}$   \\
& [erg cm s$^{-1}$] & [eV] &  [erg cm s$^{-1}$] & [eV] &  & [erg cm s$^{-1}$] &  [erg cm s$^{-1}$] \\
2000 low & \multirow{2}{*}{$9.7\pm{4.3}\times 10^{-15}$} & \multirow{2}{*}{$49^{+5}_{-4}$} & \multirow{2}{*}{$7.4^{+14.3}_{-4.7 }\times 10^{-14}$} & $-$ & $-$ &  $-$ & $ 8.4^{+14.8}_{-5.1}\times 10^{-14}$  \\
2000 high & & & & $> 250$ & $4.4^{+0.4}_{-0.8} $ & $7.8\pm{0.9}\times 10^{-13} $ & $8.6^{+2.4}_{-1.4}\times 10^{-13}$   \\
2019 low & \multirow{2}{*}{$3.3^{+0.6}_{-0.5 }\times 10^{-14}$} & \multirow{2}{*}{$48\pm{4}$} & \multirow{2}{*}{$7.9^{+14.1}_{-4.9 }\times 10^{-14}$} &  $-$ & $-$ & $-$ & $1.1^{+1.5}_{-5.1}\times 10^{-13} $  \\
2019 high & & & &$224^{+31}_{-22} $ &  $2.9\pm{0.3}$ & $7.6^{+0.4}_{-0.3}\times 10^{-13} $ & $8.8^{+1.8}_{-1.1}\times 10^{-13}$   \\
\hline
 \multicolumn{8}{c}{Model 4b: \texttt{tbabs*(pow + diskbb + comptt)}: $\chi^2/\nu=210/198$}\\
\multicolumn{8}{c}{ $N_H  = 4\pm{1}\times 10^{20}$ cm$^{-2}$  cm$^{-2}$}     \\
Spectrum  & $f_{0.3-2}^{pow}$ &  $kT^{disk}$ & $f_{0.3-2}^{disk}$ & $kT^{com}$ &  $\tau^{com}$ & $f_{0.3-2}^{com}$ & $f_{0.3-2}^{tot}$   \\
& [erg cm s$^{-1}$] & [eV] &  [erg cm s$^{-1}$] & [eV] &  & [erg cm s$^{-1}$] &  [erg cm s$^{-1}$] \\
2000 low & \multirow{2}{*}{$1.0\pm{0.4}\times 10^{-14}$} & \multirow{2}{*}{$48_{-4}^{+3}$} & \multirow{2}{*}{$7.4^{+12.6}_{-4.1}\times 10^{-14}$} & $-$ & $-$ &  $-$ & $ 8.5^{+15.5}_{-4.8}\times 10^{-14}$ \\
2000 high & & & & $440^{+2300}_{-220}$ & $7\pm{4} $ & $7.6^{+9.7}_{-7.3}\times 10^{-13} $ & $8.5^{+4.2}_{-4.5}\times 10^{-13}$   \\
2019 low & \multirow{2}{*}{$3.3^{+0.4}_{-0.3}\times 10^{-14}$} & \multirow{2}{*}{$48^{+5}_{-4}$} & \multirow{2}{*}{$7.9^{+10.6}_{-4.8}\times 10^{-14}$} &  $-$ & $-$ & $-$ & $1.1^{+1.6}_{-0.5}\times 10^{-13} $  \\
2019 high & & & &  $220_{-20}^{+30} $  &  $14^{+5}_{-2}$  & $7.6^{+3.3}_{-2.6}\times 10^{-13} $ & $8.8^{+5.5}_{-3.2}\times 10^{-13}$    \\
\hline
\hline
\end{tabular}
\end{table*}
 \end{appendix}

\end{document}